\documentclass[twocolumn,twocolappendix]{aastex631}

\usepackage{graphicx}
\usepackage{amsmath} 
\usepackage{enumerate}
\usepackage{gensymb}

\shorttitle{Latent SDEs for modeling quasar variability}
\shortauthors{Fagin et al.}

\begin{document}

\title{Latent Stochastic Differential Equations for Modeling Quasar Variability and Inferring Black Hole Properties}

\correspondingauthor{Joshua Fagin}
\email{jfagin@gradcenter.cuny.edu}

\author{Joshua~Fagin}
\affiliation{The Graduate Center of the City University of New York, 365 Fifth Avenue, New York, NY 10016, USA}
\affiliation{Department of Astrophysics, American Museum of Natural History, Central Park West and 79th Street, NY 10024-5192, USA}
\affiliation{Department of Physics and Astronomy, Lehman College of the CUNY, Bronx, NY 10468, USA}

\author{Ji~Won~Park}
\affiliation{SLAC National Accelerator Laboratory, Menlo Park, CA 94025, USA}

\author{Henry~Best}
\affiliation{The Graduate Center of the City University of New York, 365 Fifth Avenue, New York, NY 10016, USA}
\affiliation{Department of Astrophysics, American Museum of Natural History, Central Park West and 79th Street, NY 10024-5192, USA}
\affiliation{Department of Physics and Astronomy, Lehman College of the CUNY, Bronx, NY 10468, USA}

\author{James~H.~H.~Chan}
\affiliation{Department of Astrophysics, American Museum of Natural History, Central Park West and 79th Street, NY 10024-5192, USA}
\affiliation{Department of Physics and Astronomy, Lehman College of the CUNY, Bronx, NY 10468, USA}

\author{K.~E.~Saavik~Ford}
\affiliation{The Graduate Center of the City University of New York, 365 Fifth Avenue, New York, NY 10016, USA}
\affiliation{Department of Astrophysics, American Museum of Natural History, Central Park West and 79th Street, NY 10024-5192, USA}
\affiliation{Department of Science, CUNY Borough of Manhattan Community College, 199 Chambers St, New York, NY 10007, USA}
\affiliation{Flatiron Institute, 162 Fifth Avenue, New York, NY 10010, USA}

\author{Matthew~J.~Graham}
\affiliation{Flatiron Institute, 162 Fifth Avenue, New York, NY 10010, USA}
\affiliation{California Institute of Technology, 1200 E. California Blvd, Pasadena, CA 91125, USA}

\author{V.~Ashley~Villar}
\affiliation{Center for Astrophysics \textbar{} Harvard \& Smithsonian, 60 Garden Street, Cambridge, MA 02138-1516, USA}

\author{Shirley~Ho}
\affiliation{Flatiron Institute, 162 Fifth Avenue, New York, NY 10010, USA}
\affiliation{Department of Astrophysical Sciences, Princeton University, 4 Ivy Lane, Princeton, NJ 08544, USA}
\affiliation{Department of Physics and Center for Data Science, New York University, 60 5th Ave, New York, NY 10011, USA}
\affiliation{Department of Physics, Carnegie Mellon University, 10 40th St, Pittsburgh, PA 15201, USA}

\author{Matthew~O'Dowd}
\affiliation{The Graduate Center of the City University of New York, 365 Fifth Avenue, New York, NY 10016, USA}
\affiliation{Department of Astrophysics, American Museum of Natural History, Central Park West and 79th Street, NY 10024-5192, USA}
\affiliation{Department of Physics and Astronomy, Lehman College of the CUNY, Bronx, NY 10468, USA}

\begin{abstract}
Quasars are bright and unobscured active galactic nuclei (AGN) thought to be powered by the accretion of matter around supermassive black holes at the centers of galaxies. The temporal variability of a quasar’s brightness contains valuable information about its physical properties. The UV/optical variability is thought to be a stochastic
process, often represented as a damped random walk described by a stochastic differential equation (SDE). Upcoming wide-field telescopes such as the Rubin Observatory Legacy Survey of Space and Time (LSST) are expected to observe tens of millions of AGN in multiple filters over a ten year period, so there is a need for efficient and automated modeling techniques that can handle the large volume of data. Latent SDEs are machine learning models well suited for modeling quasar variability, as they can explicitly capture the underlying stochastic dynamics. In this work, we adapt latent SDEs to jointly reconstruct multivariate quasar light curves and infer their
physical properties such as the black hole mass, inclination angle, and temperature slope. Our model is trained on realistic simulations of LSST ten year quasar light curves, and we demonstrate its ability to reconstruct quasar light curves even in the presence of long seasonal gaps and irregular sampling across different bands, outperforming a multioutput Gaussian process regression baseline. Our method has the potential to provide a deeper understanding of the physical properties of quasars and is applicable to a wide range of other multivariate time series with missing
data and irregular sampling.
\end{abstract}


\section{Introduction} 

Active galactic nuclei (AGN) are among the brightest objects in the Universe and play a crucial role in galaxy evolution. They are believed to derive their immense brightness from the conversion of gravitational potential energy into thermal radiation in the hot accretion disks of supermassive black holes (SMBHs) at the center of galaxies~\citep{Salpeter64,Zeldovich64}. Luminous AGN with unobscured accretion disks are known as quasars. These objects are so luminous they remain observable at extreme cosmological distances~\citep{Mortlock11, Banados18} making them exceptional probes of the early Universe. 

The stochastic variability of quasars has been studied extensively since their discovery~\citep{1963Natur1,1963Natur2,1963ApJ,1963Natur3,1963Natur4}. While the accretion disk of a quasar is too small to be resolved at extragalactic distances, the temporal brightness variability found in quasar light curves can provide insight into the intrinsic physical properties of AGN. For instance, the variability amplitude increases with decreasing luminosity, rest-frame wavelength, and Eddington ratio~\citep{wills1993statistics, giveon1999long, berk2004ensemble}. Efforts have been made to constrain the correlation of variability parameters with black hole mass, but their robustness has been inconclusive, with studies claiming positive or negative relations depending on the degrees of observational bias present in the fit data~\citep{wold2008black, macleod2010modeling, simm2015pan}. Inferring black hole physics from quasar light curves can teach us about the evolution of the Universe and the nature of dark matter and dark energy~\citep{Khadka_2020,Czerny_2023}. 

UV/optical quasar variability is typically modeled using Gaussian process regression (GPR) by fitting light curves with a specified kernel and optimizing the kernel parameters~\citep[see][for a review of GPR in astronomical time series]{aigrain2022gaussian}. For example, \citet{Stone_2022} used the GPR package \texttt{Celerite}~\citep{celerite} to fit 20 yr quasar light curves separately in each band with a damped random walk (DRW) kernel. Markov Chain Monte Carlo (MCMC) sampling is then used to draw from the joint posterior distribution and obtain uncertainties on the variability parameters. The measured variability parameters have been empirically shown to be related to properties of the quasar such as the black hole mass~\citep{macleod2010modeling,Suberlak_2021}.

This UV/optical variability is thought to be powered by an X-ray driving variability source that illuminates the accretion disk~\citep{Cackett2007}. The reprocessing of the driving variability in the UV/optical regions of the accretion disk leads to wavelength-dependent time lags that can be measured through continuum reverberation mapping to probe the relative size scales of the emitting regions. This also makes the UV/optical variability highly correlated, since these regions are driven by the same X-ray source~\citep[e.g.,][]{Miller_2023}. The time delays between bandpasses are related to properties of the accretion disk and black hole~\citep{Cackett_2021,Jha_2022,Wang_2023}.

Upcoming wide-field surveys such as the Rubin Observatory Legacy Survey of Space and Time (LSST) mark an unprecedented improvement in data quality and volume. The LSST main survey alone (10,000~$\deg^2$) is projected to yield tens of millions of AGN light curves with six UV/optical bandpass filters (\textit{ugrizy}) at $55-185$ samplings per band, out to redshift $z \sim 7.5$ and moderate luminosities of $L \sim 10^{44}$~erg $\text{s}^{-1}$ over a ten-year period~\citep{abell2009lsst}. In the Deep Drilling Fields, a smaller sky area (200~$\deg^2$) with finer cadence, we expect to detect 40,000 additional ultrafaint AGN at about 1000 samplings per band \citep{brandt2018active}. Machine learning (ML) algorithms are well suited to analyze the large amount of data expected from LSST and other wide-field surveys; however, these light curves pose challenges for existing techniques such as multiple bands, long gaps of missing data, nonuniform sampling, and photometric and systematic noise.

Applying a GPR analysis on the entire LSST light-curve sample would be very computationally challenging, but a fully trained neural network (NN) could be rapidly deployed on all the tens of millions of AGN light curves expected from LSST. Several authors have applied ML methods to analyze quasar light curves. \citet{Tachibana_2020} used a recurrent autoencoder to model quasar variability and found it to be superior to a DRW model in real data, but it performed similarly on simulated DRW light curves. \citet{Snchez_S_ez_2021} used a recurrent variational autoencoder as a form of anomaly detection to detect changing-state AGN. \citet{Hajdinjak_2022} used conditional neural processes~\citep{garnelo2018conditional} as a nonparametric way of modeling AGN light curves. \citet{Sheng_2022} applied stochastic recurrent NNs~\citep[RNNs;][]{fraccaro2016sequential} to reconstruct light curves of simulated LSST data. In microlensed quasar light curves, \citet{Vernardos_2019} used a convolutional NN to measure the accretion disk size and temperature profile in simulations, and~\citet{Best_2022} predicted the black hole mass, inclination angle, and impact angle. \citet{JiWon_2021} introduced the first method to simultaneously reconstruct quasar light curves and predict accretion disk parameters using attentive neural processes~\citep{Kim_2019}. \citet{Danilov2022} used neural inference Gaussian processes and introduced the concept of the convolved DRW to model quasar variability, and thus outperformed standard GPR. 

In this work we adapt latent stochastic differential equations~\citep[SDEs;][]{Torch_SDE} to model irregularly sampled astronomical time series and perform parameter inference. Latent SDEs are a type of generative deep learning model that can model continuous-time stochastic dynamics. They can be viewed as infinite-dimensional variational autoencoders~\citep[VAEs;][]{VAE,rezende2014stochastic} with an SDE-induced process as their latent state. We train a latent SDE to model the dynamics of simulated ten-year LSST light curves and simultaneously predict variability and accretion disk parameters such as the timescale and strength of the variability, the black hole mass, the temperature slope, and the inclination angle. Unlike GPR on individual light curves, latent SDEs can model the variability across an entire training set without being restricted to any Gaussian process kernel and can directly infer accretion disk properties. Latent SDEs can also simultaneously fit all bandpasses at once and directly predict parameter posteriors without the need for computationally expensive MCMC sampling. 

We present our light-curve simulation in Section~\ref{sec:sim}. We introduce our latent SDE model and training in Section~\ref{sec:ML}. We describe our GPR baseline in Section~\ref{Sec:GPR}. We give our results on the performance of our model at reconstruction of light curves and inference of variability and accretion disk parameters in Section~\ref{sec:results}. Our concluding remarks and discussion are given in Section~\ref{sec:conclusion}. 

\section{Light-curve simulation} \label{sec:sim}

Our training set comprises of a realistic simulation of LSST 10 yr light curves. The UV/optical variability in our simulation is modeled as a DRW X-ray source corona situated above the black hole illuminating the accretion disk~\citep{Cackett2007,Starkey2015}. To simulate the response of the accretion disk to the driving DRW, a transfer function is calculated for each spectral band~\citep{Blandford82}. The transfer functions account for the reprocessing of the X-ray driving variability at the corona to the UV/optical variability emitted by the accretion disk, and they contain information about the quasar structure and black hole geometry.  After the light curves are simulated, they are degraded to mimic LSST-like 10 yr observation cadences and noise.

\subsection{Accretion Disk Model}

\begin{figure*}
\centering
\includegraphics[width=1.0\linewidth]{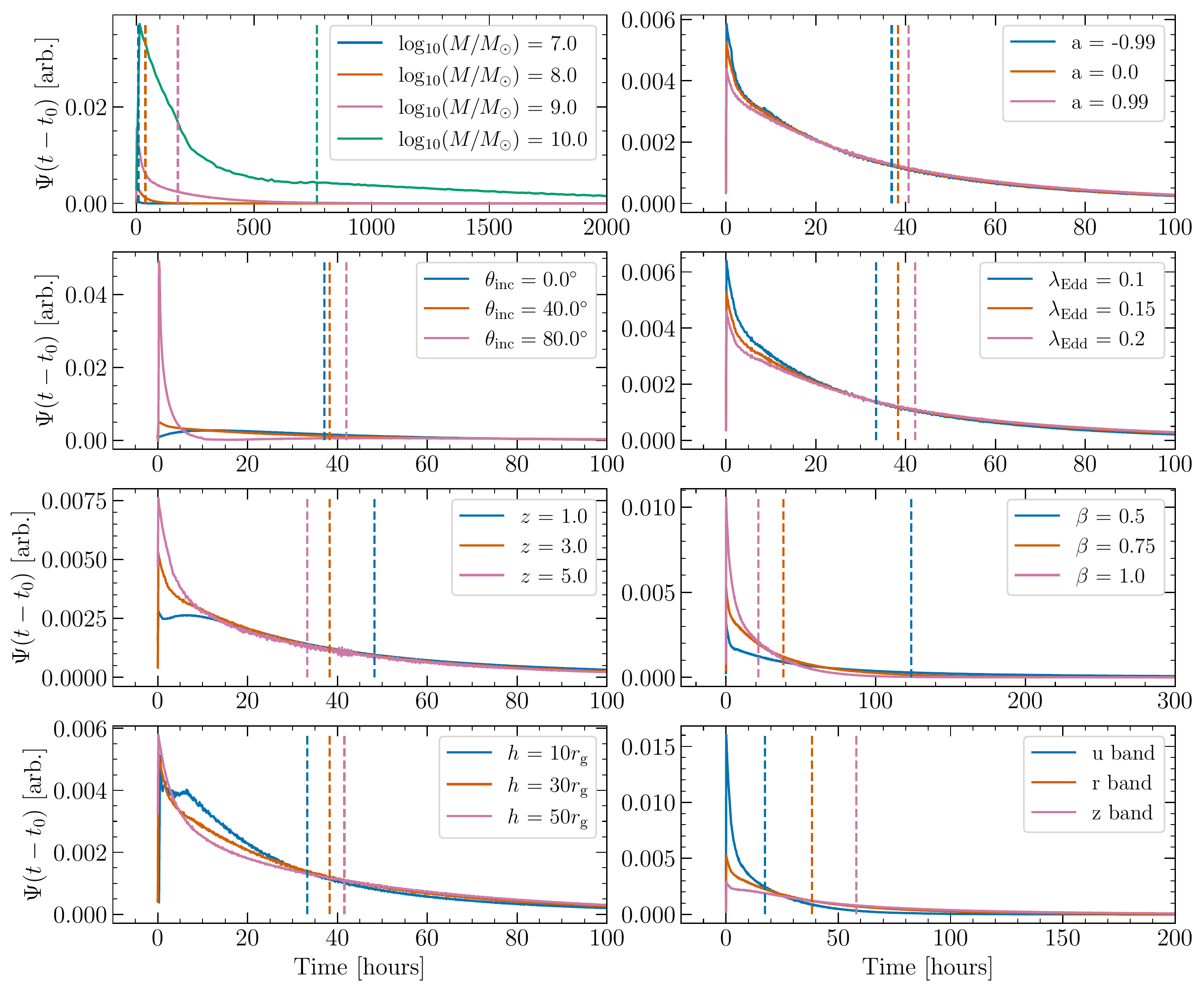}
\caption{Example of $r$-band transfer functions showing the effect of changing each parameter. We also show the impact of using different LSST bands in the bottom right panel. Any parameters not explicitly given are fixed: $\log_{10}(M/M_\odot) = 8$, $\lambda_{\text{Edd}} = 0.15$, $h = 30 r_g$, $a = 0$, $\beta = 3/4$, and $z = 3$. The mean times of the transfer functions are found using Equation~(\ref{eq:time_lag}) and shown by the vertical dashed lines (in the same color as their respective transfer functions). We introduce $t_0$ in order to get rid of the initial time lag where the transfer function would be zero.}
\label{fig:TFs}
\end{figure*}

Recent studies of quasar variability have found that quasar accretion disk sizes tend to be larger than the predictions of the thin-disk model by a factor of \mbox{$\sim$2--3} \citep{Edelson2015,Fausnaugh2016,Edelson2017,Jiang2017,Starkey2017,Kokubo_2018,McHardy2018,Mudd2018,pozo2019optical,amorim2020spatially} although this is not always the case~\citep{edelson2019first,homayouni2019sloan,hernandez2020intensive,yu2020quasar}. We use a modified version of the thin-disk model\footnote{The model will be open source in Best et al. (in preparation).} that includes special relativistic effects (Doppler beaming) and general relativistic ray tracing in a Kerr black hole geometry~\citep{Kerr_BH}. All of the physical properties of the black hole and quasar geometry are included in the transfer functions that are modeled using the general relativistic ray-traced accretion disk simulation software~\texttt{Sim5}~\citep{Sim5,Sim5_2} similar to the procedure of~\citet{Best_2022}. These transfer functions assume an X-ray driving source irradiating a vertically thin, optically thick accretion disk~\citep{ShakuraSunyaev}. The thin-disk model predicts an effective temperature profile of the form $T_{\text{eff}} \propto r^{-3/4}$. Recent studies of quasar microlensing variability studies have considered a general temperature profile of the form $T_{\text{eff}} \propto r^{-\beta}$ and favor shallower slopes $\beta < 3/4$~\citep{Cornachione_2020}. Some accretion disk models predict shallower or steeper slopes than the thin-disk slope of $\beta = 3/4$. For example, the slim disk model predicts a shallower slope of $\beta \approx 0.5$~\citep{SlimDisk} while the MRI model predicts a slope of $\beta = 7/8$~\citep{Agol_2000}. The temperature slope can also be modified due to the presence of wind outflows where the accretion rate becomes radially dependent~\citep{You_2016, Li_2018, Sun_2018}. 
We assume a modified version of the thin-disk plus lamppost temperature profile of the form
\begin{align} \label{eq:temperature}
T_{\text{eff}}^4(r) = & \frac{3GM\dot{M}}{8\pi \sigma_{\text{SB}} r_s^3} \left(1 - \sqrt{\frac{r_{\text{ISCO}}}{r}}\right) \left(\frac{r}{r_s}\right)^{-4\beta}  \\
& + \frac{(1-A)\eta_X \dot{M} c^2}{4\pi\sigma_{\text{SB}}}\frac{h}{(r^2+h^2)^{3/2}} , \nonumber
\end{align}
where $M$ is the black hole mass, $r$ is the radial distance to the SMBH, $r_s = 2r_g = 2GM/c^2$ is the Schwarzschild radius, $r_{\text{ISCO}}$ is the radius of the innermost stable circular orbit (ISCO), $\dot{M}$ is the accretion rate, $\sigma_{\text{SB}}$ is the Stefan-Boltzmann constant, $\beta$ is the temperature slope or power law of the accretion disk, $\eta_X$ is the conversion efficiency from accreted matter to X-ray radiation, and $A$ is the albedo with full absorption at $A = 0$ and full reflection at $A = 1$. The ISCO radius only depends on the SMBH spin $a$ where $r_{\rm ISCO}$ = $6r_g$ for $a = 0$ (nonrotating), $1r_g$ for $a = 1$ (maximally prograde), and $9r_g$ for $a = -1$ (maximally retrograde). \citet{Sun_2018} proposed a wind cooling parameter corresponding to a radially dependent power law in the accretion rate, which is consistent with our the first term in Equation~(\ref{eq:temperature}) in the case where $\beta < 3/4$. We generalize the power law to also consider the case when $\beta > 3/4$, consistent with~\citet{Cornachione_2020}. This could come from wind inflows or from other divergences from the thin-disk model. The second term in Equation~(\ref{eq:temperature}) is from the flux due to the lamppost corona. The Eddington ratio is defined as $\lambda_{\text{Edd}} = \dot{M}/\dot{M}_\text{Edd}$ where the Eddington accretion rate is $\dot{M}_\text{Edd} = 4\pi G M m_p/\eta\sigma_T c^2$, $m_p$ is the proton mass, $\sigma_T$ is the Thomson scattering cross section, and $\eta$ is the overall radiative efficiency factor. The transfer functions are then generated from the temperature profile using the irradiated disk model~\citep{Sergeev05, Cackett2007} at each LSST wavelength to an outer radius of 2000$r_g$ with a resolution of 0.1$r_g$. The transfer functions introduce a wavelength-dependent time lag
\begin{equation} \label{eq:time_lag}
    \bar{\tau}_{\lambda} = \frac{\int_{0}^{\infty} \Psi(t', \lambda) t' dt'}{\int_{0}^{\infty} \Psi(t', \lambda) dt'} ,
\end{equation}
into the UV/optical light curves. We normalize the transfer functions to represent a probability distribution such that $\int_{0}^{\infty} \Psi(t', \lambda) dt' = 1$ for all $\lambda$. 

Examples of transfer functions are shown in Figure~\ref{fig:TFs}, where different sets of parameters are varied. In this figure we set the initial time lags to zero to better compare the impacts of different parameters. For example, a greater corona height $h$ leads to a longer initial time lag because photons need to travel a longer distance from the corona to the disk. Each band will have the same initial time lag, however, so the lag does not impact our parameter predictions. It is clear from the figure that varying the mass has the largest effect on the transfer functions. The temperature slope also has a significant impact on the mean time delay of the transfer function. The inclination angle has little impact on the mean of the transfer functions but significantly changes its shape. The redshift, height, and Eddington ratio have more minor effects, while the black hole spin has a negligible impact. 

An example of the difference in the mean time delay between the $r$ and $u$ bands demonstrating the impact the parameters have on the mean time delay between the two bands is shown in Appendix~\ref{sec:appendix_parameters}. Since many parameters impact the mean time delays in similar ways, there are degeneracies that make predicting the accretion disk parameters challenging. There is also a degeneracy between the transfer functions and the X-ray driving variability. Convolving the driving signal with the transfer function kernel eliminates some of the high-frequency variations, making the new time series appear to have a longer variability timescale. This means our parameter predictions must be primarily driven by the time delays between bands, with higher-order effects arising from the differences in the variability timescales across bands. 

\subsection{Model of the Quasar Driving Variability}

\begin{figure*}[ht!]
\centering
\includegraphics[width=0.9\textwidth]{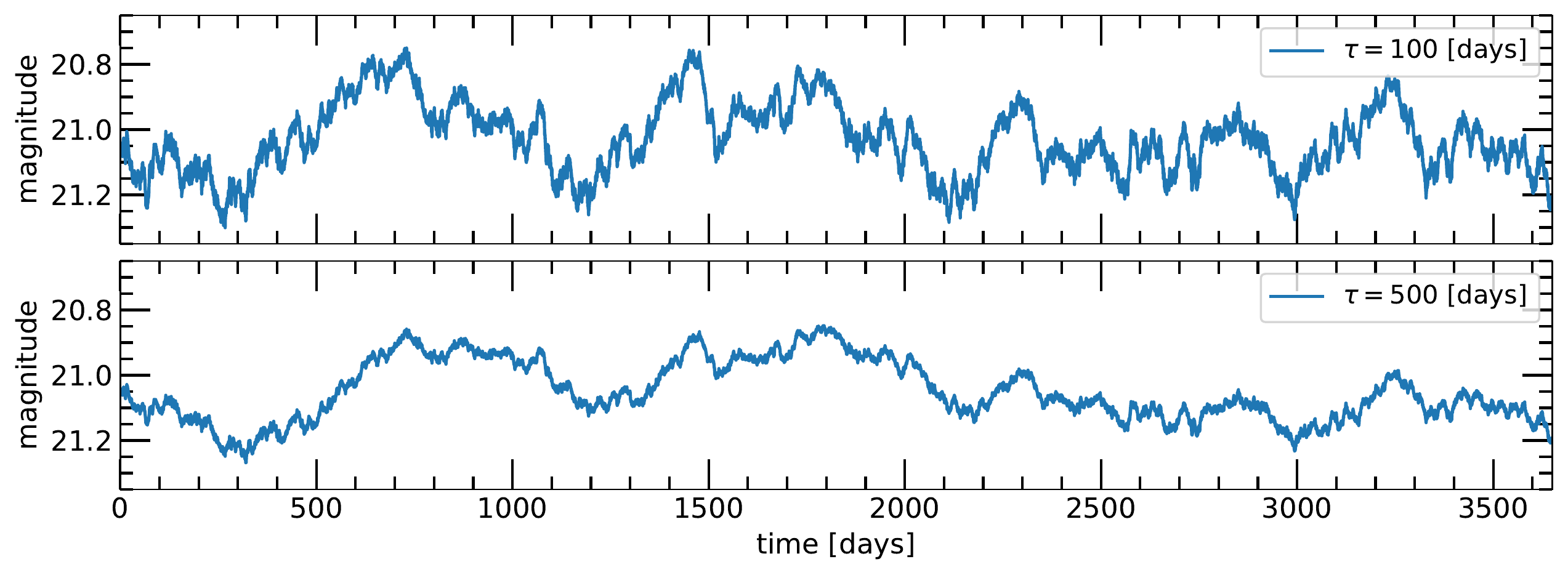}
\caption{Damped random walk for two different values of $\tau$ generated using Equation~(\ref{eq:DRW}) with \mbox{$\overline{X}=21$~mag}, \mbox{$\text{SF}_\infty=0.1$~mag}, and $z = 0$. For comparison, both DRWs use the same white noise process so that their stochastic dynamics are identical.}
\label{fig:DRW}
\end{figure*}
 
Quasar UV/optical variability is often modeled as a damped random walk, a type of Gaussian process also known as the Ornstein-Uhlenbeck process~\citep{williams2006gaussian,Zu_2013}. A DRW signal $X(t)$ is governed by the SDE:
\begin{equation} \label{eq:DRW}
dX(t) = -\frac{1}{\tau} X(t)\ dt +\sigma \sqrt{dt}\ \epsilon(t) + b\ dt
\end{equation}
where $\epsilon(t)$ is a white noise process with mean zero and variance one, $\tau$ is the characteristic timescale, $b$ is related to the mean of the process $\overline{X}=b\tau$, and $\sigma$ is related to the standard deviation of $X(t)$ defined by the asymptotic structure function $\text{SF}_\infty = \sigma\sqrt{\tau/2}$~\citep{Kelly2009}. 
GPR can be used to measure the variability parameters $\text{SF}_\infty$ and $\tau$, which may be correlated with the black hole mass~\citep{macleod2010modeling,Suberlak_2021}. A smaller $\tau$ yields more high-frequency variations since the variability happens on a shorter timescale. Figure~\ref{fig:DRW} shows examples of DRWs for two values of $\tau$. 

The kernel describing the covariance between two observations separated in time by $\Delta t$ is given by 
\begin{align} \label{eq:gp_kernel}
    k(\Delta t) & = \text{SF}_\infty^2 \cdot \exp\left(-\frac{\Delta t}{\tau} \right) \\ 
    & = \text{SF}_\infty^2 \cdot \text{ACF}(\Delta t) , \nonumber
\end{align}
where the ACF is the autocorrelation function. An alternative description of the process is the structure function (SF), which is the rms scatter of magnitude differences $\Delta m$ as a function of the temporal separation $\Delta t$. For the DRW process, the SF takes the form
\begin{align} \label{eq:sf}
    {\rm SF}(\Delta t) & = {\rm SF_\infty} \sqrt{ 1 -  \exp\left(-\frac{\Delta t}{\tau} \right) } \\ 
    & = {\rm SF_\infty} \sqrt{1-\text{ACF}(\Delta t)} . \nonumber 
\end{align}
The DRW is a red noise process with power spectral density (PSD) given by
\begin{equation}
P(f) = \frac{4\tau\text{SF}_\infty^2}{1+(2\pi \tau f)^2} ,
\end{equation}
which implies that $P(f) \propto f^{-2}$ at high frequencies when $f \gg (2\pi \tau)^{-1}$ and $P(f)$ is constant at lower frequencies. 

If the local variability of the corona follows a DRW with characteristic timescale $\tau_{\text{rest}}$, then we will observe a boosted variability with $\tau_{\text{obs}} = (1+z)\tau_{\text{rest}}$. We boost our DRW to the observer frame when simulating light curves, but we train our network to predict $\tau_{\text{rest}}$ since it is the timescale that governs the intrinsic variability of the quasar and should be correlated to the physical properties of the accretion disk (for example, the black hole mass). Hereafter, $\tau$ refers to the variability timescale in the rest frame.

We model the X-ray corona driving variability using the DRW process. Although X-ray variability may be better described by a more general broken power-law PSD~\citep{McHardy_2004,Oneill_2005,Uttley_2005,soton466440,Markowitz_2010,Yang_2022,yuk2023correlation}, we use the simple DRW prescription to try to recover its parameters $\tau$ and $\text{SF}_\infty$ using latent SDEs and so that we can use an appropriate kernel for our GPR baseline. The DRW is itself a special case of the bending broken power law with a lower power of $0$ and higher power of $-2$. We postulate that latent SDEs would only show improved performance compared to GPR with more complex driving variability due to their flexibility and GPR not having a suitable kernel. 

Our X-ray driving DRW is generated in magnitude to avoid the possibility of getting negative values for the flux~\citep{Czerny_2023}. We select a mean magnitude independently of any of the variability or black hole parameters, drawing from the distribution $\overline{X} = \mathcal{N}(20.90,1.04)$ mag. The mean and standard deviation of this distribution come from the average magnitude of the LSST bands across the Sloan Digital Sky Survey and Hyper Suprime-Cam survey~\citep{Ahumada_2020,Chan_2024}. We choose this strategy to avoid biasing our parameter estimations with priors on the correlation between absolute magnitude and recovered parameters including $z$, $\lambda_{\rm Edd}$, and $\log_{10}(M/M_{\odot})$. With actual LSST data, however, where any correlation is the direct result of physical processes we would aim to understand, the network could be fine-tuned with the data to include these processes in its training, and we would expect the parameter recovery to improve. The DRW driving variability is converted from AB magnitude to flux\footnote{$m_{\text{AB}} = -2.5 \log_{10}(f_\nu)-48.60$ in cgs units.} and then convolved with the transfer function kernels to obtain simulated UV/optical light curves. The light curves are then converted from flux back to magnitude. 

Our simulated UV/optical light curves are not strictly DRWs because they result from the convolution of the driving DRW with the transfer functions. The convolution smooths out the driving variability, and can change the power spectrum of the light curve by eliminating some of the high-frequency variations. For the majority of cases, the timescale of the transfer functions is significantly less than the timescales of the variability $\tau$, making the effect minor. The convolution also introduces a time delay between bands caused by the difference in the mean time of the transfer functions. These time delays are related to the accretion disk parameters that produced the transfer functions. The mean time lags introduced by the transfer functions range from less than a day for less massive black holes and up to more than a month for the most massive black holes (see the top left panel of Figure~\ref{fig:TFs}).

\subsection{Parameter Ranges}

\begin{table}
\begin{center}
\caption{Parameters and ranges for our light-curve simulation. The parameters are sampled uniformly between their minimum and maximum. We also fix the radiative efficiency to $\eta = 0.1$ and X-ray radiative efficiency and albedo to $(1-A)\eta_X = 0.005$} 
 \begin{tabular}{l l c c} 
Name & Description & Min. & Max. \\
 \hline\hline
$\text{SF}_\infty$/mag & DRW amplitude & $0.0$ & $0.5$ \\
$\log_{10}(\tau/\text{day})$ & DRW timescale & $-0.5$ & $3.5$ \\
$\log_{10}(M/M_\odot)$ & BH mass & $7.0$ & $10.0$ \\
$a$ & BH spin & $-1.0$ & $1.0$ \\
$\theta_{\rm inc}$ & Inclination angle & $0.0^{\circ}$ & $80.0^{\circ}$ \\
$h/r_g$ & Corona height & $10.0$ & $50.0$ \\
$\beta$ & Temp. power law & $0.3$ & $1.0$ \\
$z$ & Redshift & $0.1$ & $6.0$ \\
$\lambda_{\text{Edd}}$ & Eddington ratio & $0.01$ & $0.3$ \\
 \hline 
\end{tabular}
\label{table:Parameter_ranges}
\end{center}
\end{table}

Our light-curve simulation is determined by a mean magnitude and nine physical parameters given in Table~\ref{table:Parameter_ranges}. The driving X-ray variability is dependent on $\text{SF}_\infty$ and $\tau$. The black hole geometry is determined by its mass $M$ and dimensionless spin parameter $a$. The quasar is at an inclination angle $\theta_{\text{inc}}$ with respect to the observer and at a redshift $z$. We also vary the height of the X-ray corona $h$, the temperature slope of the accretion disk $\beta$, and the Eddington ratio $\lambda_{\text{Edd}}$. There is also the radiative efficiency that we fix to the commonly used value of $\eta = 0.1$~\citep[e.g.,][]{Fausnaugh_2018}. This radiative efficiency is degenerate with the Eddington ratio, so our measurement of the Eddington ratio can be thought of as $\lambda_{\text{Edd}}\eta/0.1$. We also fix the X-ray radiative efficiency and albedo to a typical value of $\eta_X(1-A) = 0.005$, since $\eta_X = \eta/\lambda_{\text{Edd}} \cdot L_X/L_{\text{Edd}}$ with $L_X/L_{\text{Edd}} \sim 0.005$~\citep{Ursini_2020} and the albedo $A$ is typically small, i.e., $A\sim$0.1--0.2~\citep{Qiao_2018}. We fix these values to attempt to measure the corona height with a consistent corona strength. We use the effective wavelengths of each LSST band: $\lambda_{\text{eff}, u} = 3671$\AA, $\lambda_{\text{eff}, g} = 4827$\AA, $\lambda_{\text{eff}, r} = 6223$\AA, $\lambda_{\text{eff}, i} = 7546$\AA, $\lambda_{\text{eff}, z} = 8691$\AA, $\lambda_{\text{eff}, y} = 9712$\AA 
~\citep{Huber_2021} to produce the transfer functions.

\subsection{Mock LSST Observations}

After a UV/optical light curve is simulated, we degrade it to impose LSST-like observing cadences and noise. The total error of an LSST observation is given by
\begin{equation} \label{eq:noise}
\sigma_{\text{LSST}}^2 = \sigma_{\text{sys}}^2 + \sigma_{\text{rand}}^2 ,
\end{equation}
where $\sigma_{\text{sys}}$ is the systematic error and $\sigma_{\text{rand}}$ is the photometric noise. We set the systematic error to $0.005$ mag, which is the maximum value expected for LSST~\citep{LSST,Suberlak_2021}. The photometric noise is expected to follow
\begin{gather}
\sigma_{\text{rand}}^2 = (0.04-\gamma)x+\gamma x^2 \quad (\text{mag}^2) , \\
x = 10^{0.4(m-m_5)} \nonumber
\end{gather}
where $\gamma$ is a band-dependent factor, $m$ is the magnitude at each observation, and $m_5$ is the $5\sigma$ depth of a point source at the zenith of each visit. We set $\gamma_{u} = 0.38$ and $\gamma_{g, r, i, z, y} = 0.39$~\citep{LSST,Sheng_2022}. We impose LSST-like observation cadences and noise by using \texttt{rubin\_sim}\footnote{\href{https://github.com/lsst/rubin\_sim}{https://github.com/lsst/rubin\_sim}} with the \texttt{baseline\_v2.1\_10yrs} rolling cadence to obtain the time and $m_5$ of each observation. We sample 100,000 points anywhere in the sky that have between 750 and 1000 total observations across the ten years to include only light curves from the main LSST survey \citep[see Figure 1 of][]{Prsa_2023}. The inclusion of light curves from the Deep Drilling Fields that have higher cadences would improve the performance of our model. Once the LSST data become available, the network could be fine-tuned on the data or retrain using the exact cadence strategy as part of our simulation.

The error added to each observation is randomly sampled from a Gaussian with mean zero and variance $\sigma_{\text{LSST}}^2$. Observations are combined to the nearest one-day interval, averaging across $\sigma_{\text{rand}}^2$ if there were multiple observations in a band in the same interval. We simulate 11 yr light curves with 10 yr of LSST observations so our model is trained to reconstruct some time before and after the survey. We randomly select the start time of the first LSST observation within the first year of the 11 yr light curves. 

\section{Latent SDEs}  \label{sec:ML}

In this section we describe our latent SDE model and training. The model and code are open source and available at: \href{https://github.com/JFagin/latent_SDE}{https://github.com/JFagin/latent\_SDE}.

\subsection{Model Architecture}

\begin{figure}
\centering
\includegraphics[width=1.0\linewidth]{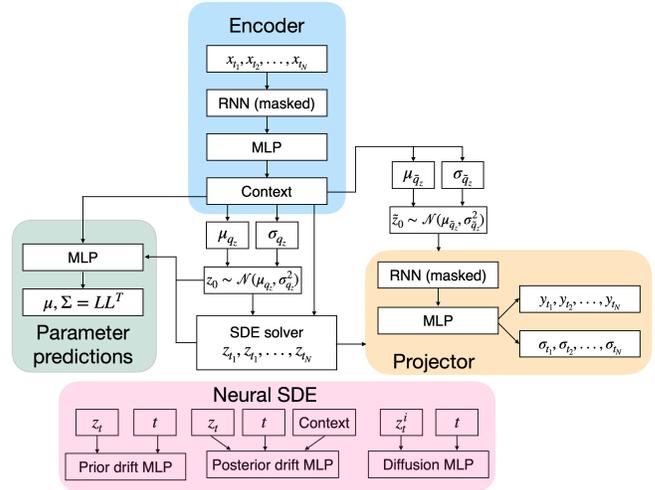}
\caption{Diagram of the latent SDE architecture.}
\label{fig:model}
\end{figure}

The input to the network is the brightness and error values (both in magnitude), for a total of 12 features at each time step (6 bands and 6 errors). For each band that is not observed at a given time step, we set both the brightness and error to a dummy value of zero to be masked by the network. The only preprocessing step required to apply the network to LSST data is to combine observations to the nearest daily interval and set the unobserved time steps to zero.

The network architecture can be broken down into three main parts: the RNN encoder, the neural SDE decoder, and the multilayer perceptron (MLP) parameter estimator. Figure~\ref{fig:model} provides a high-level overview of the model. Overall, the network operates by first encoding the multivariate time series into a context using an RNN that includes masking of the unobserved times. Then the mean and standard deviation of the variational distribution are predicted from the context. A latent vector is drawn from the variational distribution and then the latent vector and context are used by the neural SDE solver to produce a time series with features equal to the latent size. That time series as well as the errors of the input light curve are then projected using another RNN into the mean and standard deviation of our reconstructed light curve. Another latent vector is also employed by the projector to improve the uncertainty estimation on our reconstructed light curve. In addition to the light-curve reconstruction, we predict the mean and covariance matrix for the parameter inference based on the context and latent state of our SDE. We jointly model the light curve and parameters since the light-curve dynamics should be directly related to the variability and accretion disk parameters.

The NN architecture uses a hidden size of 128 and a context size of 64. A latent size of 8 is used for the SDE with an additional latent vector of size 8 given to the RNN projector to aid in uncertainty estimation. Each hidden layer is followed by a LeakyReLU activation function~\citep{maas2013rectifier} and layer normalization~\citep{layer_norm}. 
The network is built using \texttt{PyTorch}~\citep{Pytorch} and the SDE solver uses the \texttt{torchsde}~\footnote{\href{https://github.com/google-research/torchsde}{https://github.com/google-research/torchsde}}~\citep{Torch_SDE} implementation. Our network has a total of 903,597 trainable parameters.
The light curves are normalized to have mean zero and variance one during training (only using observed points to calculate the mean and variance and averaged across the bands). 

The encoder starts with a GRU-D~\citep{GRUD} layer, a modified version of the gated recurrent unit~\citep[GRU;][]{GRU} that is designed to handle both masking and irregular time intervals. We use the GRU-D to mask the unobserved portions of our light curves. The GRU-D layer is followed by two GRU layers and two fully connected layers that produce the context. We include a residual skip connection~\citep{ResNet} between the GRU-D layer and the output of the second GRU layer, which is intended to improve gradient flow and training stability. Two more fully connected layers are used to produce the mean $\mu_{q_z}$ and variance $\sigma_{q_z}^2$ of the variational distribution $q_z$ from the last value, mean, and standard deviation of the context and the mean and standard deviation of the unnormalized light curve. We then draw the latent vector from the variational distribution $z_0 \sim \mathcal{N}(\mu_{q_z},\,\sigma_{q_z}^{2})$, the initial state of the SDE.

The decoder includes an It{\^o} SDE solver with diagonal noise containing networks governing the drift, diffusion, and prior drift~\citep{Torch_SDE}. We use the Euler–Maruyama numerical approximation scheme for the SDE solver. The drift and prior drift networks are both MLPs with a residual skip connection. There are three fully connected layers pre-skip and two post-skip. The diffusion network consists of three fully connected layers and is applied element-wise to satisfy the diagonal noise. In order to model both the dynamics of the time series as well as produce uncertainties on predictions, we use an RNN to project the output of the SDE and the input errors of the light curve onto the mean and log-variance of the observation space. The projector also includes a new latent vector $\tilde{z}_0 \sim \mathcal{N}(\mu_{\tilde{q}_z},\,\sigma_{\tilde{q}_z}^{2})$ that is repeated across each time step in the input. We include the additional latent vector to enhance the uncertainty estimation since the latent SDE and its latent vector should model the dynamics of the time series, but we also want to model the uncertainty at each time step. The projector consists of a GRU-D layer followed by two GRU layers. There is also a skip connection of two fully connected layers between the output of the SDE and the output of the RNN.

The parameter inference component of our model uses the same MLP architecture as the drift networks and outputs the mean and Cholesky factor of the covariance matrix for the multivariate Gaussian posterior (described formally in Section~\ref{sec:loss}). The input into the parameter estimator is the last value, mean, and standard deviation of the context, $\mu_{q_z}$, $\sigma_{q_z}^2$, the output of the drift, diffusion, and prior drift networks at $z_0$, and the mean and standard deviation of the unnormalized light curve. 

\subsection{Uncertainty Quantification and Loss Function} \label{sec:loss}

For light-curve reconstruction, we minimize the negative evidence lower bound (ELBO) described in \cite{Torch_SDE}. We use a negative Gaussian log-likelihood between the reconstructed and true light curve averaged across the bands:
\begin{equation}
\mathcal{L} = \frac{1}{N} \sum_{t=1}^N \frac{(y_t-\hat{y}_t)^2}{2\hat{\sigma}_t^2} + \frac{1}{2}\log(2\pi\hat{\sigma}_t^2) , 
\label{NGLL}
\end{equation}
where $y$ is the true light curve, $\hat{y}$ is the predicted mean, $\hat{\sigma}^2$ is the predicted variance, and $N$ is the number of time steps. In practice we predict $\log(\hat{\sigma}^2)$ to force the variance to be positive. To ensure that the mean prediction closely matches the context points, we additionally evaluate the weighted mean squared error of the mean prediction with respect to the context points and their errors, averaged across the bands:
\begin{equation} 
\mathcal{L} = \frac{1}{N_{\text{obs}}} \sum_{t=1}^N m_t \cdot \frac{(y_{\text{LSST},t}-\hat{y}_t)^2}{2\sigma_{\text{LSST},t}^2} ,
\label{eq:context_loss}
\end{equation}
where $y_{\text{LSST}}$ is the mean observation or context point with variance $\sigma_{\text{LSST}}^2$, $m_t$ is a mask that is 1 if a band is observed at a time step $t$ and 0 if it is not, and \mbox{$N_{\text{obs}} = \sum_{t=1}^N m_t$} is the number of observations. We find that this additional term in the loss function leads to much better performance by overemphasizing the context points with respect to the mean prediction.

For the parameter inference, we parameterize the posterior as a multivariate Gaussian distribution and minimize its negative log-likelihood:
\begin{equation}
\mathcal{L} = \frac{1}{2} (\boldsymbol{y}-\boldsymbol{\hat{y}})^\top \Sigma^{-1} (\boldsymbol{y}-\boldsymbol{\hat{y}}) + \frac{1}{2} \log(2\pi|\Sigma|) ,
\label{eq:NGLL_multi}
\end{equation}
where each $\boldsymbol{y}$ is the vector of true parameter values, $\boldsymbol{\hat{y}}$ is the vector of predicted means, and $\Sigma$ is the predicted covariance matrix. In practice we predict the lower triangular matrix $L$ representing the Cholesky decomposition of the covariance matrix $\Sigma = LL^\top$ to ensure that the covariance matrix is symmetric, positive semidefinite, and nonsingular. We must also restrict the diagonal of $L$ to be positive so the covariance matrix is positive semidefinite and the parameterization is unique, so we predict the log-diagonal of $L$ and then take the exponential. To predict $n$ parameters, the lower triangular matrix $L$ will have \mbox{$n(n+1)/2$} free parameters. Each term in the loss is given equal weighting.

When simulating the light curves, each parameter is drawn uniformly from a given range. We reparameterize the labels from their physical values to between zero and one and then take the logit (scaling the labels from $-\infty$ to $\infty$) before evaluating the negative log-likelihood of our posterior. Once samples are drawn from the posterior, we take the sigmoid (forcing the predictions to be between zero and one again) and then scale them back to the original physical range. This prevents posterior probability from being wasted in physically impossible parameter space (such as restricting the spin to \mbox{$-1 < a < 1$} for example) or outside the range of the training set. 

\subsection{Training}

We train our network with 100,000 light curves per epoch that are randomly regenerated on the fly each epoch to prevent overfitting and improve performance. We also include a fixed validation set and a test set of 10,000 light curves each. Each set uses separate transfer functions to avoid contamination. 

Our network is trained for 100 epochs using the Adam optimizer~\citep{Adam} with an initial learning rate of 0.0025, an exponential decay of 0.97, and a batch size of 50. To mitigate the problem of exploding gradients, we employ gradient clipping with a maximum gradient norm of 0.5. Linear KL-annealing~\citep{fu2019cyclical} is used for the first 15 epochs as well as annealing for the parameter inference portion of the loss (Equation~(\ref{eq:NGLL_multi})) to initially focus on the light-curve reconstruction. Training took approximately six weeks on a single NVIDIA Tesla V100 GPU with 16 GB of VRAM.

\section{Gaussian process regression baseline} \label{Sec:GPR}

We aim to compare the performance of latent SDEs in light-curve reconstruction with an exact, multitask GPR baseline~\citep{MultiTaskGPR}. Our multitask GPR model can infer correlations between output variables, in our case the different LSST bandpass filters, unlike with \texttt{Celerite}~\citep{celerite}, which fits a separate GP to each band. While some researchers, such as~\citet{Hu_2020}, have explored using GPR with multiband light curves, the more conventional approach is to fit separate GPs. To fairly compare GPR to our latent SDE model, we must jointly model bands so that observations in one band carry information across the other bands. This is crucial since each band is related to the same X-ray driving DRW process, and LSST observations are sparse and distributed asynchronously across the different bands.

The noise variance at each observation is set from the LSST observation strategy (see Equation~(\ref{eq:noise})) so we use the fixed-noise GPR model. 
We use the absolute exponential kernel (equivalent to the Matérn-1/2 kernel), which corresponds to the DRW process (i.e., Equation~(\ref{eq:gp_kernel})). This kernel has also been empirically shown to fit UV/optical quasar variability better than the Matérn-3/2, Matérn-5/2, rational quadratic, and squared exponential kernels~\citep{Griffiths_2021}. When fitting the GP, each band of the light curve is independently normalized to have mean zero and variance one, using only observed points and taking into account the error on each context point. Our fixed-noise multitask GPR baseline is implemented in \texttt{BoTorch}\footnote{\href{https://botorch.org/}{https://botorch.org/}}~\citep{BoTorch}, based on the GP library \texttt{GPyTorch}~\citep{Gpytorch}. It uses the intrinsic co-regionalization model~\citep[ICM;][]{NIPS2007_66368270,NIPS2013_f33ba15e} to correlate the kernels of different tasks (i.e., the different LSST bands in our case).

\begin{figure*}[ht!]
\centering
\includegraphics[width=1.0\textwidth]{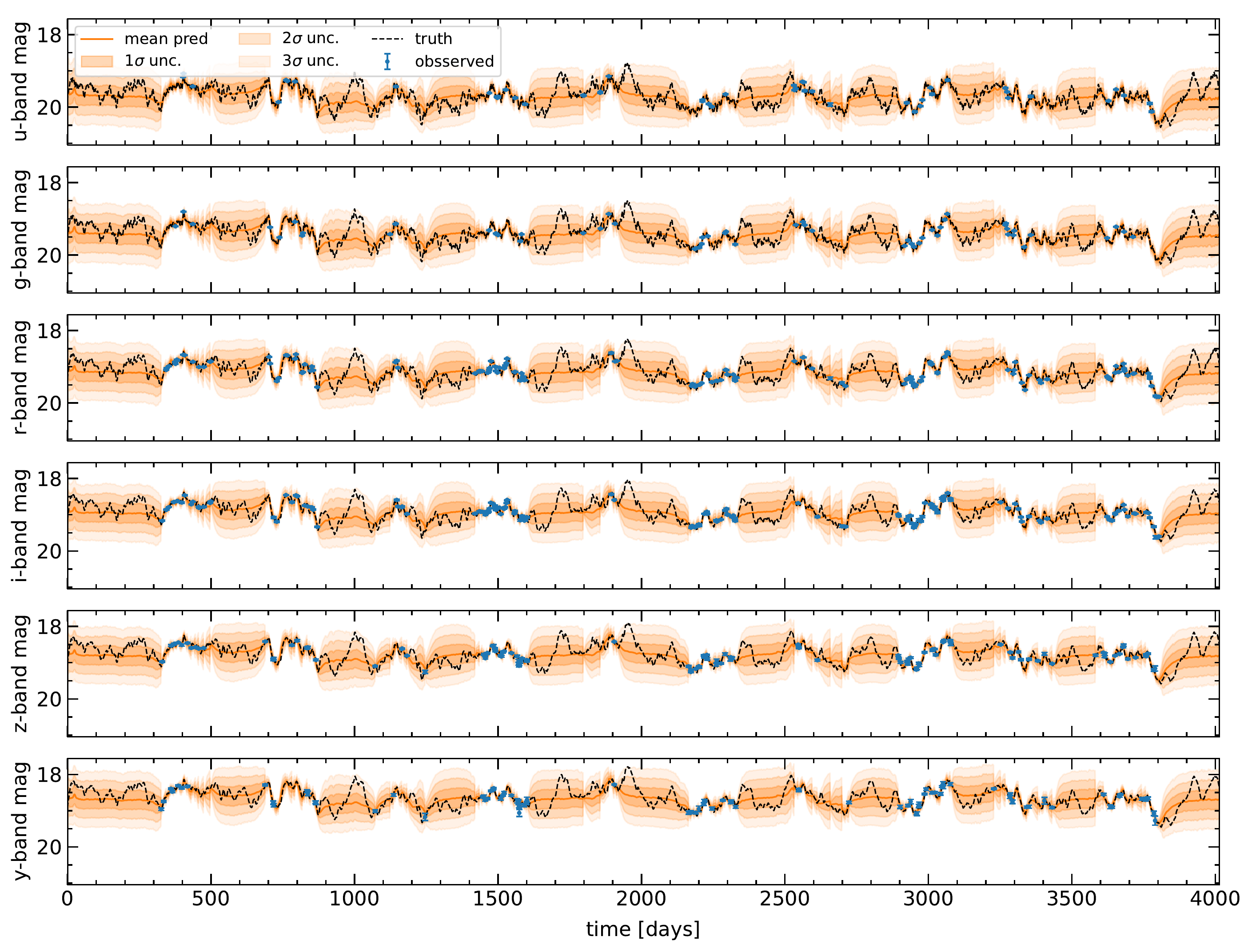}
\caption{Example of a simulated 11 yr light curve (black) from our test set with 10 yr of mock LSST observations or context points (blue) and our latent SDE reconstruction (orange) including at unobserved times.
}
\label{fig:recovery}
\end{figure*}

\section{Results} \label{sec:results}

\subsection{Light-curve Reconstruction Performance}

\begin{figure}
\centering
\includegraphics[width=1.0\linewidth]{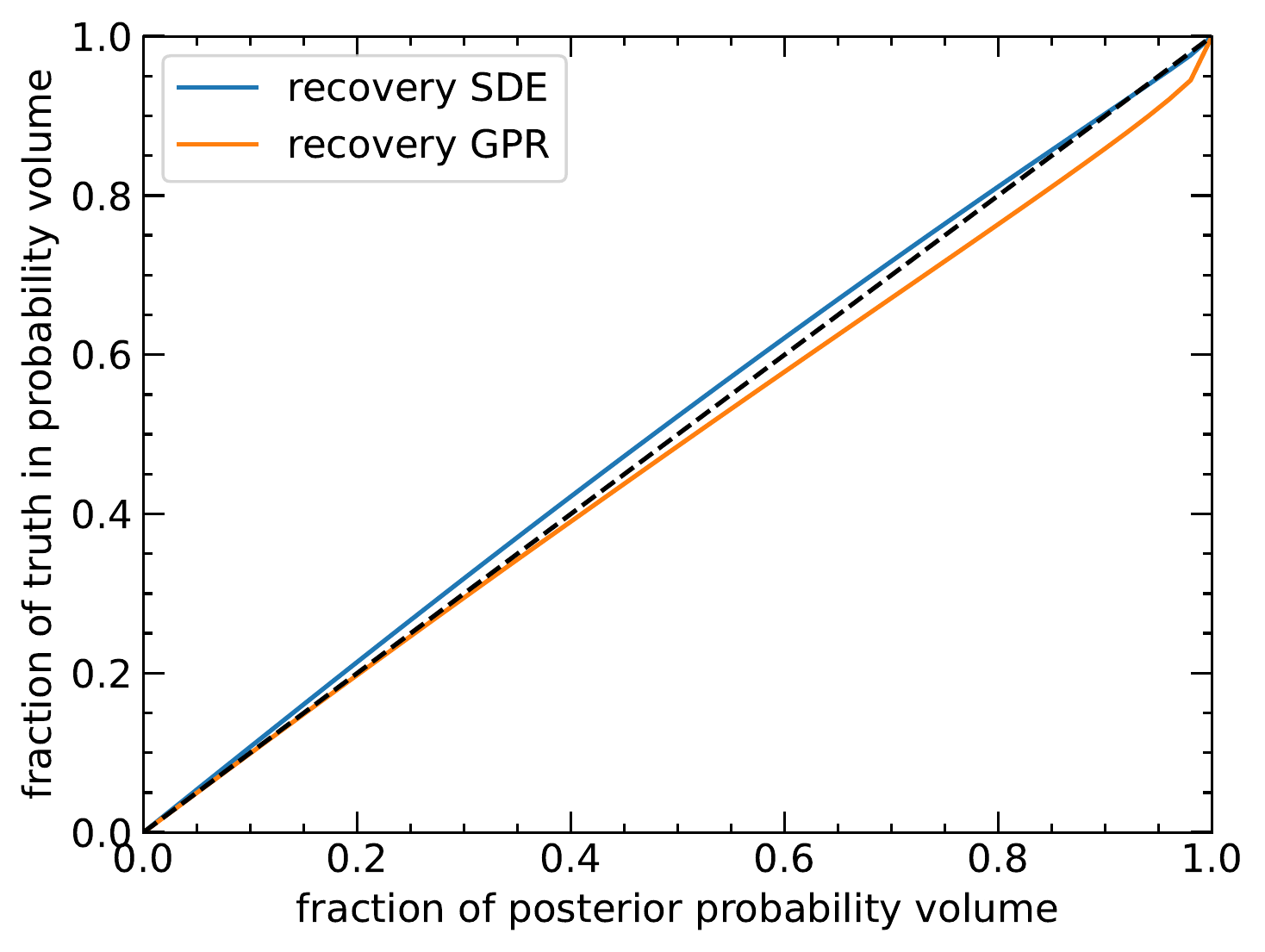}
\caption{The fraction of the truth encompassed by some probability volume (credible interval) across our test set for our latent SDE model's light-curve reconstruction (blue) compared to the GPR baseline (orange). Perfect calibration is shown by the black dashed line along the diagonal. 
}
\label{fig:coverage_prob_rec}
\end{figure}

After our model is trained, we apply it to the test set of light curves that are produced from driving variability and transfer functions that the model has never trained on. Figure~\ref{fig:recovery} shows the light-curve reconstruction for an example in the test set. As expected, the model is more uncertain at times farther away from the context points such as during the seasonal gaps. Our network learns to tightly constrain the light curve across all bands given a context point in only a single band. This is because in our training set, the variability in each band of the light curve is derived from the same driving variability source and only differs based on the transfer function kernels. For comparison, the reconstruction of the same light curve with our multitask GPR baseline is shown in Appendix~\ref{sec:appendix_GPR} and is similarly able to infer correlations between bands.

Table~\ref{table:latent_SDE_vs_GPR} compares the light-curve reconstruction performance of our latent SDE model with the GPR baseline across the test set. The latent SDE model performs better than the GPR baseline on light-curve reconstruction in terms of the rms error (RMSE), mean absolute error (MAE), and negative Gaussian log-likelihood (NGLL). In order to assess the significance of these differences, we conduct a paired $t$-test yielding extremely low $p$-values across all metrics ($<10^{-10}$). This gives very high confidence that the differences in performance between the two models are not a product of random variation but can be attributed to the latent SDE model's capabilities compared to the GPR baseline.

We also show how well the uncertainty is calibrated for our light-curve reconstruction in Fig~\ref{fig:coverage_prob_rec}. The GPR baseline is overconfident in its predictions, so the true light curve can sometimes fall outside the $3\sigma$ credible region. The latent SDE model is a little underconfident, so its predicted uncertainty may be slightly overestimated.
An example of structure function recovery is given in Appendix~\ref{sec:appendix_SF} between the true light curve, the SF for the X-ray driving DRW, the mean latent SDE prediction, and the mean GPR prediction for the same example of a light curve shown in Figure~\ref{fig:recovery}. 

\begin{table}
\begin{center}
\caption{Light-curve reconstruction performance of our latent SDE model compared to a multioutput GPR baseline. Values reported are the median $\pm$ median absolute deviation on the median for each metric across our test set of 10,000 light curves. Lower is better for all the metrics.}
\begin{tabular}{c c c}
Metric & latent SDE & GPR \\
\hline
RMSE (mag) & $0.0959\pm 0.0006$ & $0.0978\pm 0.0006$ \\
MAE (mag) & $0.0695\pm 0.0004$ & $0.0711\pm 0.0004$ \\
NGLL & $-1.14\pm 0.006$ & $-1.01\pm 0.006$ \\
\end{tabular}
\label{table:latent_SDE_vs_GPR}
\end{center}
\end{table}

\subsection{Parameter Estimation Performance}

\begin{figure*}
\centering
\includegraphics[width=1.0\textwidth]{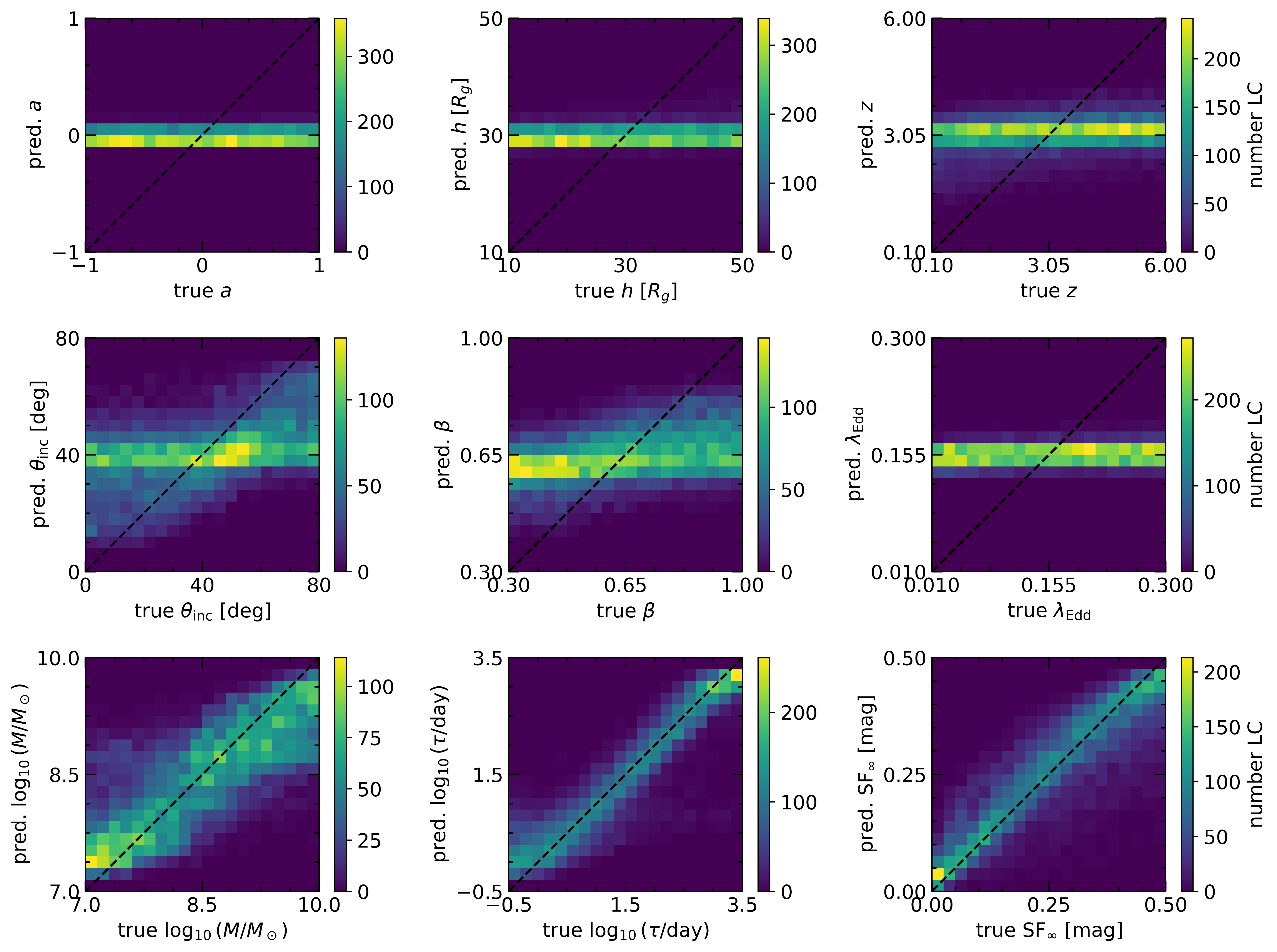}
\caption{The mean predictions compared to the true values for each parameter across our test set. The black dashed lines along the diagonal represent perfect predictions.
}
\label{fig:mean_vs_truth}
\end{figure*}

\begin{figure}
\centering

\includegraphics[width=1.0\linewidth]{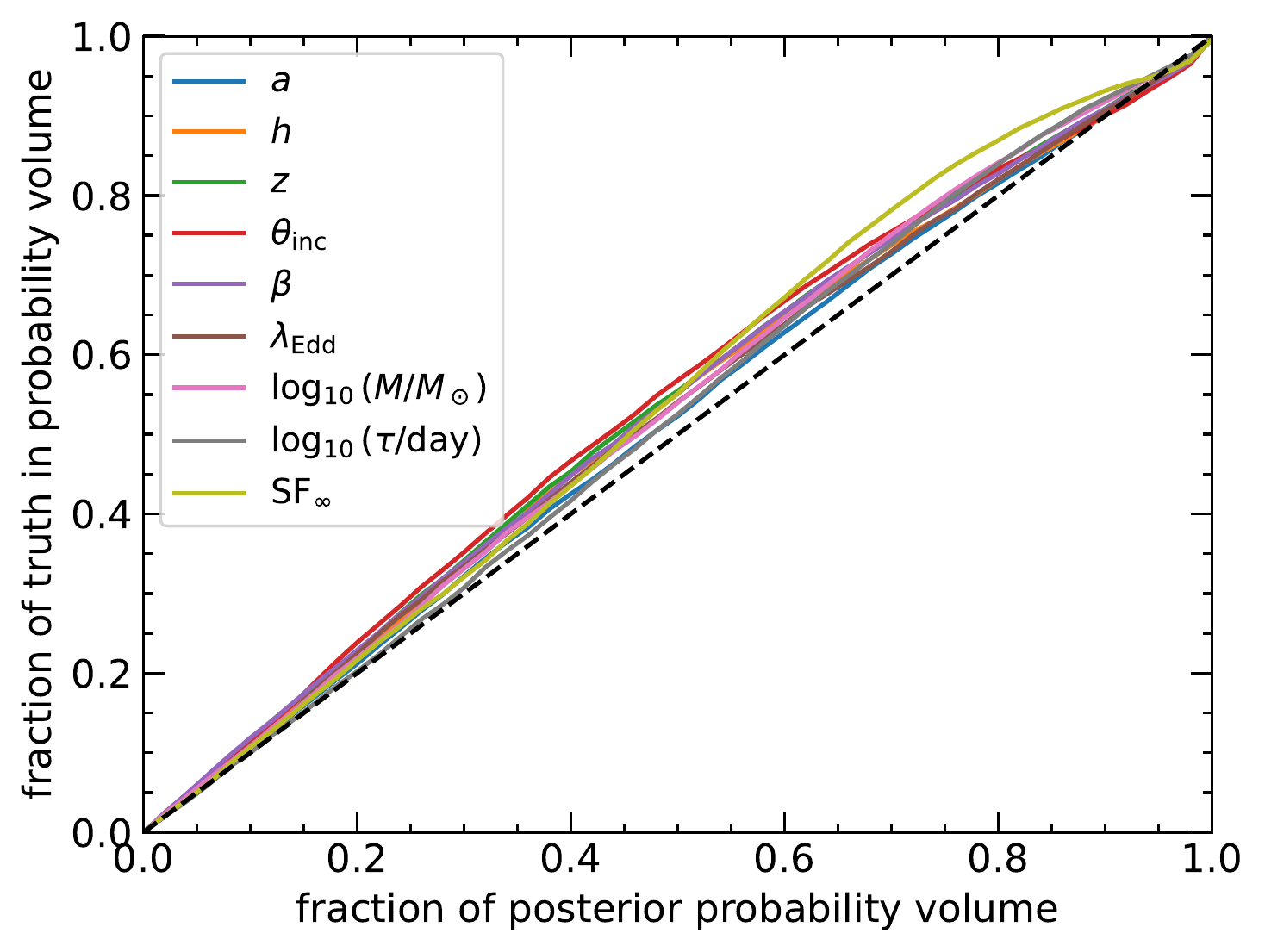}
\caption{The fraction of the truth encompassed by some probability volume (credible interval) across our test set for our parameter inference. Perfect calibration is shown by the black dashed line along the diagonal. 
}
\label{fig:coverage_prob}
\end{figure}

In Figure~\ref{fig:mean_vs_truth} we compare the network's mean parameter predictions with their true values. The network is able to predict the variability parameters $\log_{10}(\tau)$ and $\mathrm{SF}_\infty$ well across a broad parameter range. We recover the rest-frame $\tau$ well despite having poor constraints on the redshift. This is because the unknown redshift will be incorporated into the uncertainties of the predictions without a large change in the mean, and we predict $\log_{10}(\tau)$ while the redshift only affects $\tau$ in a linear way. The black hole mass is harder to predict since it only affects the transfer functions that are convolved with the DRW and must be extrapolated from the time delays between bands. The inclination angle $\theta_{\text{inc}}$ and temperature slope $\beta$ have smaller effects on the mean and shape of the transfer function than the black hole mass, but we can still make meaningful predictions in many cases. This is especially the case for the higher-mass black holes where the transfer functions produce time delays at timescales of at least a few days. The network only has a little predictive power for the redshift $z$, Eddington ratio $\lambda_{\text{Edd}}$, black hole spin $a$, and corona height $h$. This is to be expected since these parameters have only minor impact on the mean time delays between bands, as illustrated in Appendix~\ref{sec:appendix_parameters}. In the case where the network is completely unsure about the value of a parameter, it will predict mean values near the mean of the prior with large uncertainties. We show 100 examples of predictions with uncertainties in Appendix~\ref{sec:appendix_scatter}. Cases where the median predictions are far from the true value have larger uncertainties to account for the deviation.
Figure~\ref{fig:coverage_prob} shows that our reported uncertainties of the parameter posteriors are, on average, correctly calibrated across the test set, although we are slightly underconfident in our predictions. This likely arises from the limited predictive strength of certain parameters, especially for smaller black hole masses. In such situations, our parameterization of the posterior as a Gaussian would not be ideal since there is equal probability the parameters are anywhere in our prior. 

An example of the predicted posterior of our parameters is shown in Appendix~\ref{sec:appendix_param_posterior}. Even though the uncertainties on individual quasar light curves are often large, with a large sample of predictions like the tens of millions of quasar light curves expected from LSST, it is possible to make precise estimates of the population-level distributions using hierarchical inference. Hierarchical inference can combine the parameter posteriors of individual light curves into population-level predictions given that the priors on our parameter space are known and build into the training set.

\section{Conclusion and discussion}  \label{sec:conclusion}

We have presented latent SDEs as a way of simultaneously modeling quasar variability and predicting black hole properties. A pretrained latent SDE could be quickly and autonomously applied to the entire LSST quasar light-curve sample, whereas using GPR with MCMC sampling to obtain posteriors on variability parameters may be computationally infeasible. Training and inference with our model can be easily scaled with batch support across multiple GPUs, while using batched GPU support for GPR is an open area of research and not supported by \texttt{Celerite} for example. Our network is trained to infer the variability parameters of the quasar X-ray source corona while GPR can only infer the effective parameters after the driving variability is convolved with the transfer functions. In addition, GPR is limited to optimizing the GP's kernel parameters, whereas latent SDEs have the flexibility to infer any variability parameterization and reconstruct the light curve in a nonparametric way, bypassing the need for an analytic kernel. Typically GPR can only be applied separately on different bands, but here we used a multitask GPR baseline that is capable of correlating the GPs across the different bands. Our deep learning approach outperformed the state-of-the-art GPR baseline at reconstructing LSST-like quasar light curves. 

The method presented here is completely general and can be applied to any multivariate time series with missing data and irregular sampling, with applications to other astronomical time series and beyond. In our use case, the latent SDE model learns to tightly constrain the light curve across bands using a context point in a single band due to the nature of our training set. We would expect our latent SDE model to be applicable to multivariate time series where the features (i.e. the bands in our case) are less correlated, although a larger latent size may be required to capture the additional complexity. In such a
situation, the network would no longer be able to use context points in one band to tightly constrain the other bands. If the features were completely uncorrelated, it may be preferable to model each feature with a separate latent SDE instead of jointly. Parameter inference could still be performed from the joint latent space.

There is also the potential to improve the latent SDE model. Our model uses an RNN encoder and latent SDE decoder including a GRU-D layer~\citep{GRUD} to mask the unobserved times. In this work we use unidirectional RNN layers, which process time series sequentially in the forward direction. Using bidirectional RNN layers, such that the time series is encoded and decoded both forwards and backwards in time, may improve the performance of our model but would increase its computational complexity. We expect this to also smooth out the uncertainty in the reverse direction in areas where there are limited context points, such as around 2700 days in Figure~\ref{fig:recovery}. We may also explore other RNN architectures such as stochastic RNNs, used to model quasar variability in~\citet{Sheng_2022}. Recently~\citet{schirmer2022modeling} introduced the continuous recurrent unit, which is a probabilistic recurrent architecture that includes a hidden state that evolves as a linear stochastic differential equation. We may also consider using other architectures instead of RNNs, such as transformers~\citep{vaswani2023attention}. For example, multi-time attention networks~\citep[mTANDs;][]{shukla2021multitime} feed the time embedding to the attention mechanism to incorporate irregularly sampled time series. Transformers typically have many more parameters than RNNs, however, so training them in a latent SDE network may be difficult. Improvements to our latent SDE model will be investigated in future work.

Further developments in GPR frameworks could improve its performance on quasar light-curve reconstruction. The multitask GPR model does not explicitly account for the time delays between bands introduced by the convolution of the driving signal with the transfer functions. Latent SDEs learn to model the time delays since they are present in the training set. Latent SDEs are also able to jointly model accretion disk parameters such as the mass, inclination angle, and temperature slope from the light curve. These parameters are related to the time delays between bands and their relative smoothness, but no comparison currently exists with GPR.
Promising areas for advancing GPR performance include the neural inference of Gaussian processes, as applied to quasar variability in~\citet{Danilov2022}, and the exploration of deep Gaussian processes~\citep{damianou2013deep}.

Another area of application of our latent SDE model that was not explored in this work is for anomaly detection. The latent space of our model may be a powerful tool to detect quasars with anomalous variability behavior. \citet{Snchez_S_ez_2021} used the latent space of a recurrent VAE to search for changing-state AGN, and our model could be used similarly. There is also the potential to identify periodic signals from SMBH binaries or to search for transient events. 

In future work, we could improve our quasar light-curve simulation to make it more realistic. Our simulated light curves were generated with a DRW X-ray source variability convolved with physics-based transfer functions that reprocess the driving variability into the UV/optical light curves. In reality, the X-ray variability is not exactly a DRW process but may be better described by the more general broken power-law PSD~\citep{McHardy_2004,Oneill_2005,Uttley_2005,soton466440,Markowitz_2010,Yang_2022,yuk2023correlation}. We use the DRW model so that we can test our model's ability to recover its parameters and to properly use the corresponding kernel for our GPR baseline. Our latent SDE model could be easily retrained or fine-tuned with more complicated variability models such as higher-order continuous-time autoregressive moving-average (CARMA) processes~\citep{Kelly_2014,Stone_2022,Yu_2022} or generated from a general PSD using the method of~\citet{Timmer_1995}. A general PSD could come from fitting X-ray variability data or be modeled by a broken power law. We expect the performance gap between latent SDEs and GPR to grow with more complex variability because latent SDEs are more flexible in situations in which no Gaussian process kernel exists. \citet{Tachibana_2020}, for example, found that a recurrent autoencoder performed similarly to the DRW GP for simulated DRW light curves, but the deep learning approach outperformed the GP model on real data when the variability is more complex.

In our simulation, we assumed the variability parameters were uncorrelated to our black hole properties, but there are measured correlations between the variability parameters and the black hole mass, for example, although their robustness has been inconclusive~\citep{wold2008black, macleod2010modeling, simm2015pan}. We do not take these correlations as a prior in our simulation so that the correlations could be measured in real data without bias. If we did choose to model the correlation between the variability and accretion disk parameters, we would expect to improve the performance of the parameter predictions, since measuring the variability parameters would then help to constrain the redshift, Eddington ratio, and black hole mass. 

Our simulation also did not model the additional information present in the relative fluxes of each bandpass, which could be used to infer the photometric redshift. A full treatment of the color information would require modeling the quasar spectrum and integrating transfer functions across the response function of each broadband filter. In particular, the relative fluxes of each band convey information about the quasar's redshift, as the redshift determines which spectral lines are captured by each broadband filter. By modeling the quasar spectrum and brightnesses, we anticipate a significant improvement in redshift estimation, and a minor enhancement in estimating other parameters due to the better constrained redshift. Care would need to be taken to avoid biasing predictions from inconsistencies between simulated spectra and real data.

LSST will observe too many quasars to obtain follow-up spectra to make precise redshift measurements for all the monitored quasars. A subset of the LSST quasar sample will get spectral measurements, and the precise redshift measurements could be used to enhance the parameter inference. There may also be mass measurements from the broad-line region~\citep{Panda_2019} and these measurements could be used to better constrain the mass and other accretion disk parameters. 

Once the LSST data become available, we could fine-tune our network in a self-supervised way using only the context points of the light curves (minimizing the loss function given in Equation~(\ref{eq:context_loss}) for example). We could also use supervised learning with a subset of the LSST sample, with training labels coming from spectral redshift, black hole mass estimates from the broad-line region, or other parameters that have been independently measured through a variety of ways. This could partially address the issue in ML when a network is trained with simulations that do not exactly match the real data. In particular, the network could learn the complex relationships between the variability and accretion disk parameters, the quasar brightness, and the photometric redshifts that we did not include in the training set but will be present in the real data. These relationships would then be a result of physical processes and not artificially imposed as priors in the training set, and we can expect our parameter estimates to improve. A limited number of quasars will be monitored in the Deep Drilling Fields, and our network is well suited to take advantage of the much higher sampling rate to better constrain the quasar parameters. Fine-tuning our network with the higher cadences may be especially useful. 

There is also the possibility that even for parameters that are difficult to predict for individual quasar light curves, hierarchical inference could be used to estimate the population-level distributions of the parameter space. For example, measuring the population-level distribution of the temperature slope could constrain accretion disk and wind outflow models. 

\section*{Acknowledgments}
This research was made possible by the generosity of Eric and Wendy Schmidt by recommendation of the Schmidt Futures program. Matthew Graham acknowledges support from National Science Foundation (NSF) AST-2108402. V. Ashley Villar acknowledges support from NSF under grant AST-2108676. The data used in this publication were collected through the MENDEL high performance computing (HPC) cluster at the American Museum of Natural History. This HPC cluster was developed with National Science Foundation (NSF) Campus Cyberinfrastructure support through Award \#1925590. The authors would like to thank Favio Neira for assistance with \texttt{rubin\_sim}. We also thank Xuechen Li and David Duvenaud for their thoughtful comments on our work.

\software{\texttt{Matplotlib}~\citep{Matplotlib}, \texttt{Numpy}~\citep{Numpy}, 
\texttt{scipy}~\citep{Scipy}, 
\texttt{Astropy}~\citep{Astropy}, 
\texttt{astroML}~\citep{astroML}, 
\texttt{corner.py}~\citep{corner}, 
\texttt{BoTorch}~\citep{BoTorch},
\texttt{torchsde}~\citep{Torch_SDE}, 
\texttt{PyTorch}~\citep{Pytorch}}

\appendix

\section{Example of the Impact of Parameters on Mean Times of the Transfer Functions} \label{sec:appendix_parameters}

\begin{figure*}[ht!]
\centering
\includegraphics[width=0.95\textwidth]{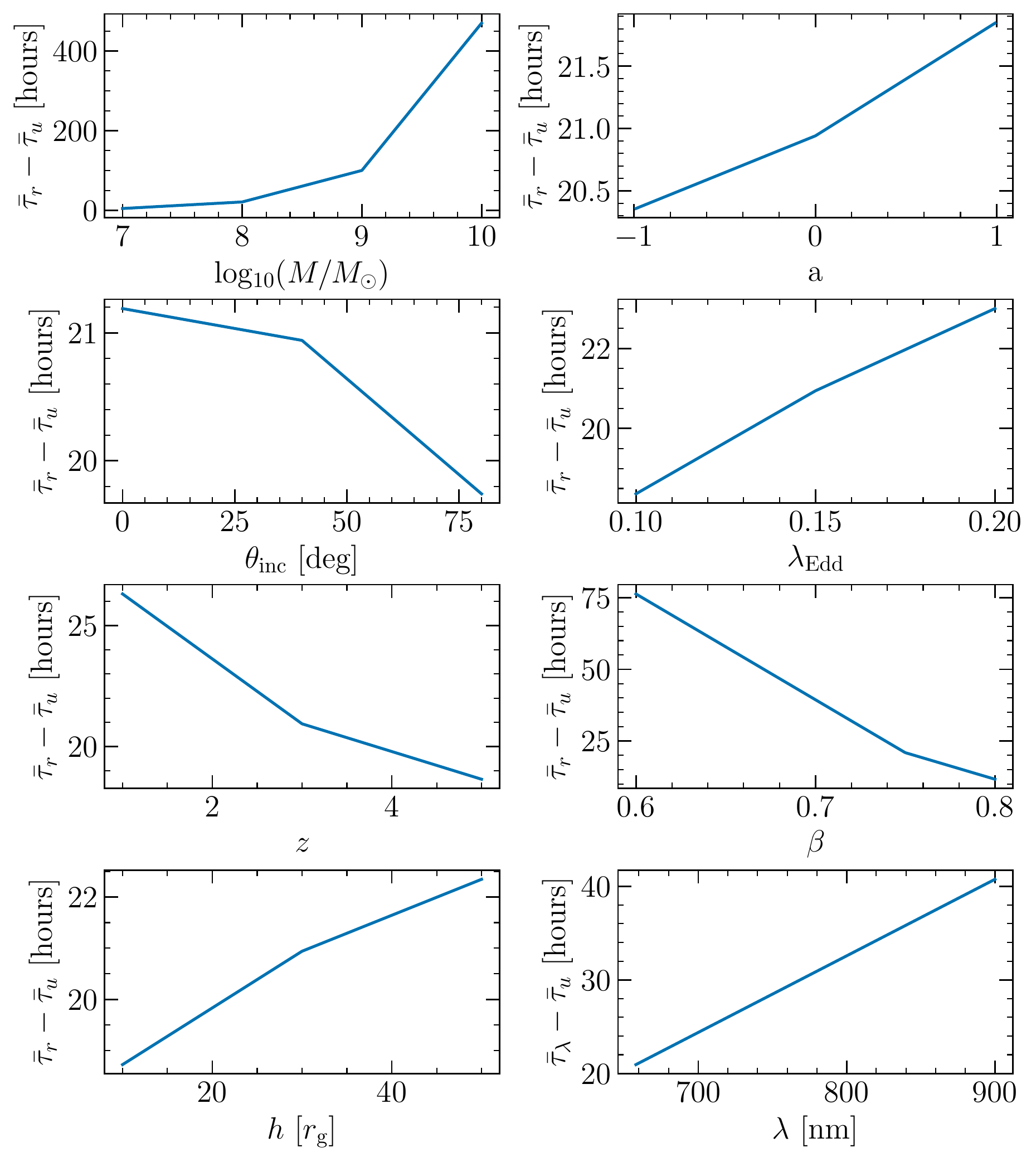}
\caption{The difference in the mean time lags between the $r$-band and $u$-band transfer functions for different parameters. We also show the impact that a change in the wavelength has on the time delay in the bottom right panel. Any parameters not explicitly given are fixed: $\log_{10}(M/M_\odot) = 8$, $\lambda_{\text{Edd}} = 0.15$, $h = 30 r_g$, $a = 0$, $\beta = 3/4$, and $z = 3$. Figure~\ref{fig:TFs} shows the $r$-band transfer functions for the same parameter range.}
\label{fig:TF_mean}
\end{figure*}

The main effect the transfer functions have on UV/optical variability is by introducing wavelength-dependent time delays (see Equation~(\ref{eq:time_lag})). We can only observe the relative time delays between bands because a constant time delay across all wavelengths is degenerate with the start time of the driving variability. Figure~\ref{fig:TF_mean} shows an example of the difference in the mean times between the $r$-band and the $u$-band transfer functions. This demonstrates the impact each parameter has on the differences in time delay between bands and approximately how well we can expect to measure the parameters. Parameters that only change the mean time delays by less than half the time binning of our time series ($< 12$ hr) my be infeasible to recover. There is an additional effect in that the transfer functions will also have a wavelength dependence in their standard deviation, making the variability of higher wavelength bands smoother. The inclination angle, for example, has very little impact on the mean time delay, but has a large effect on the width of the transfer function so can still be constrained in many cases (see Figure~\ref{fig:TFs} to see the transfer functions for different $\theta_{\text{inc}}$).

\section{Example of Light-Curve Reconstruction Using Gaussian Process Regression} \label{sec:appendix_GPR}

\begin{figure*}
\centering
\includegraphics[width=1.0\textwidth]{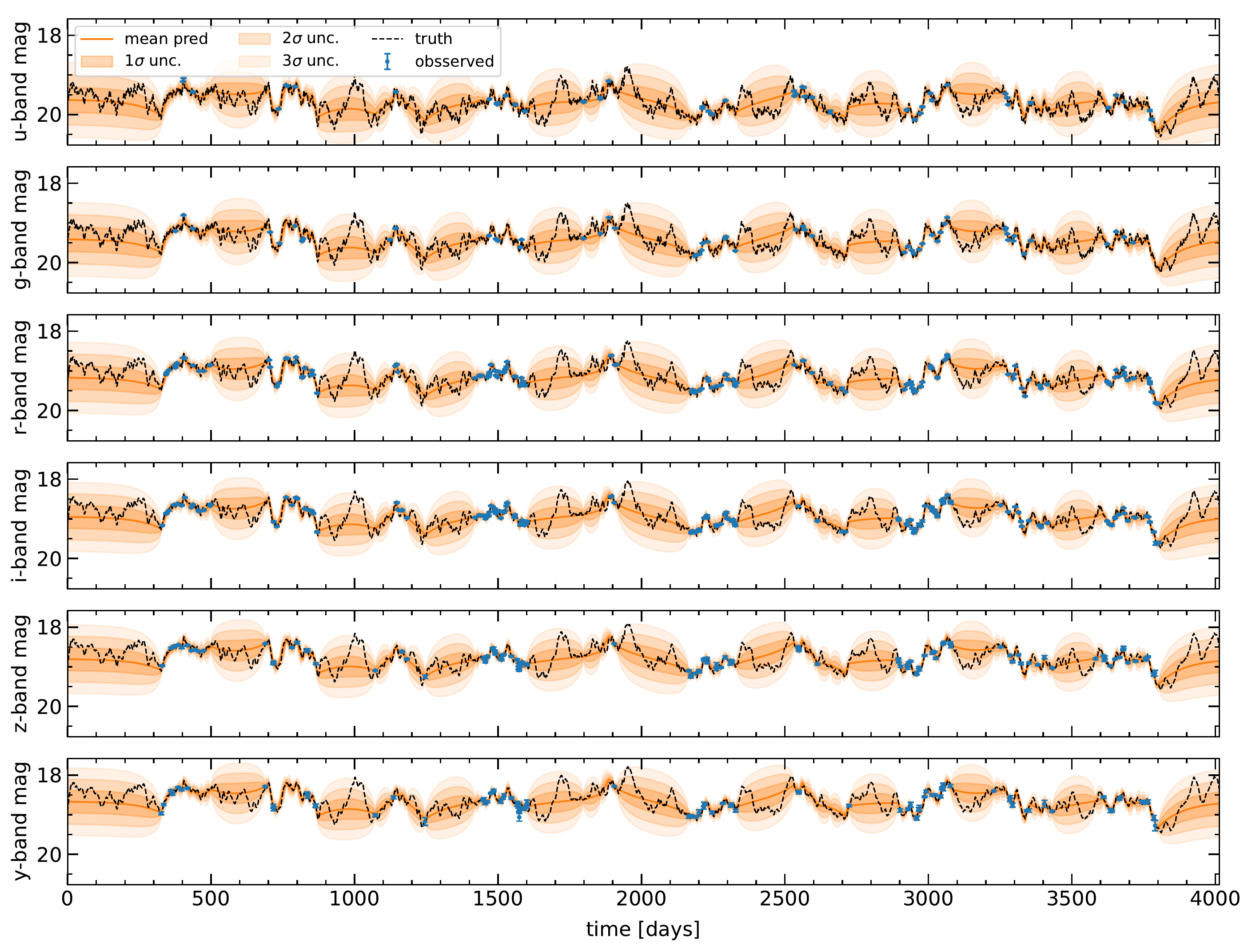}
\caption{Same as Figure~\ref{fig:recovery} but using the GPR baseline described in Section~\ref{Sec:GPR}.}
\label{fig:GPR_recovery}
\end{figure*}

Figure~\ref{fig:GPR_recovery} shows an example of light-curve reconstruction using our GPR baseline.

\section{Example of Structure Function Recovery} \label{sec:appendix_SF}

\begin{figure*}
\centering
\includegraphics[width=1.0\textwidth]{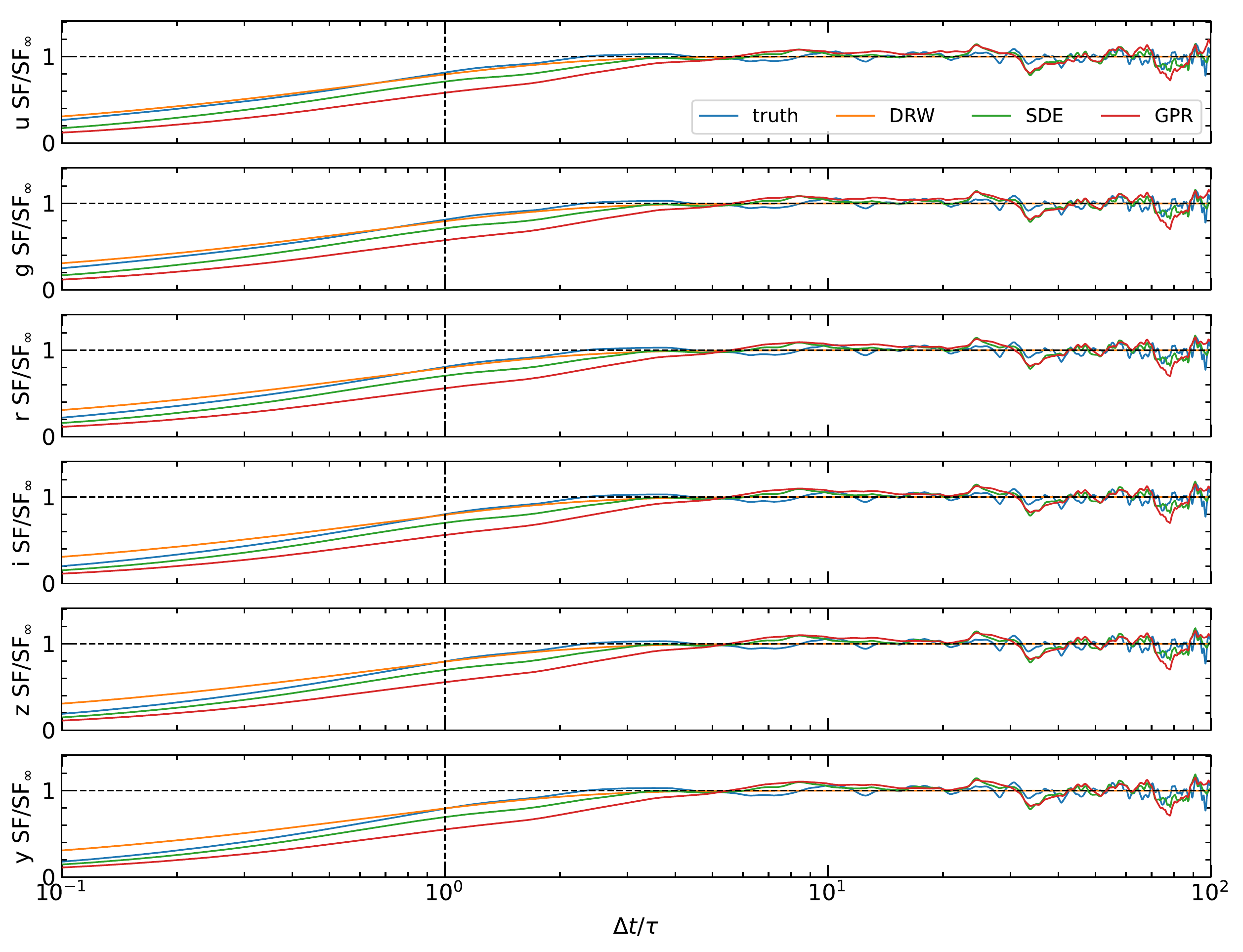}
\caption{Example of recovered structure function for the same test light curve shown in Figure~\ref{fig:recovery} using the mean prediction of our latent SDE (green) and GPR baseline (red). The true SF (blue) comes from using the light curve before any observational effects. The DRW SF (orange) is the SF corresponding to the X-ray driving variability given in Equation~(\ref{eq:sf}), but will diverge from the truth due to the convolution of the driving variability with the transfer functions. The horizontal dashed line shows when $\text{SF} = \text{SF}_\infty$ and the vertical dashed line when $\Delta t = \tau$}
\label{fig:SF}
\end{figure*}

In Figure~\ref{fig:SF} we show an example structure of function recovery using the mean recovery of our latent SDE and GPR baseline. We could also sample different predictions from the latent space of our latent SDE to get some uncertainty on the SF recovery. The mean recoveries of both the latent SDE and GPR mean recoveries generally follow the true SF.

\section{Example of Scatter in Mean Parameter Inference with Uncertainty} \label{sec:appendix_scatter}

\begin{figure*}
\centering
\includegraphics[width=1.0\textwidth]{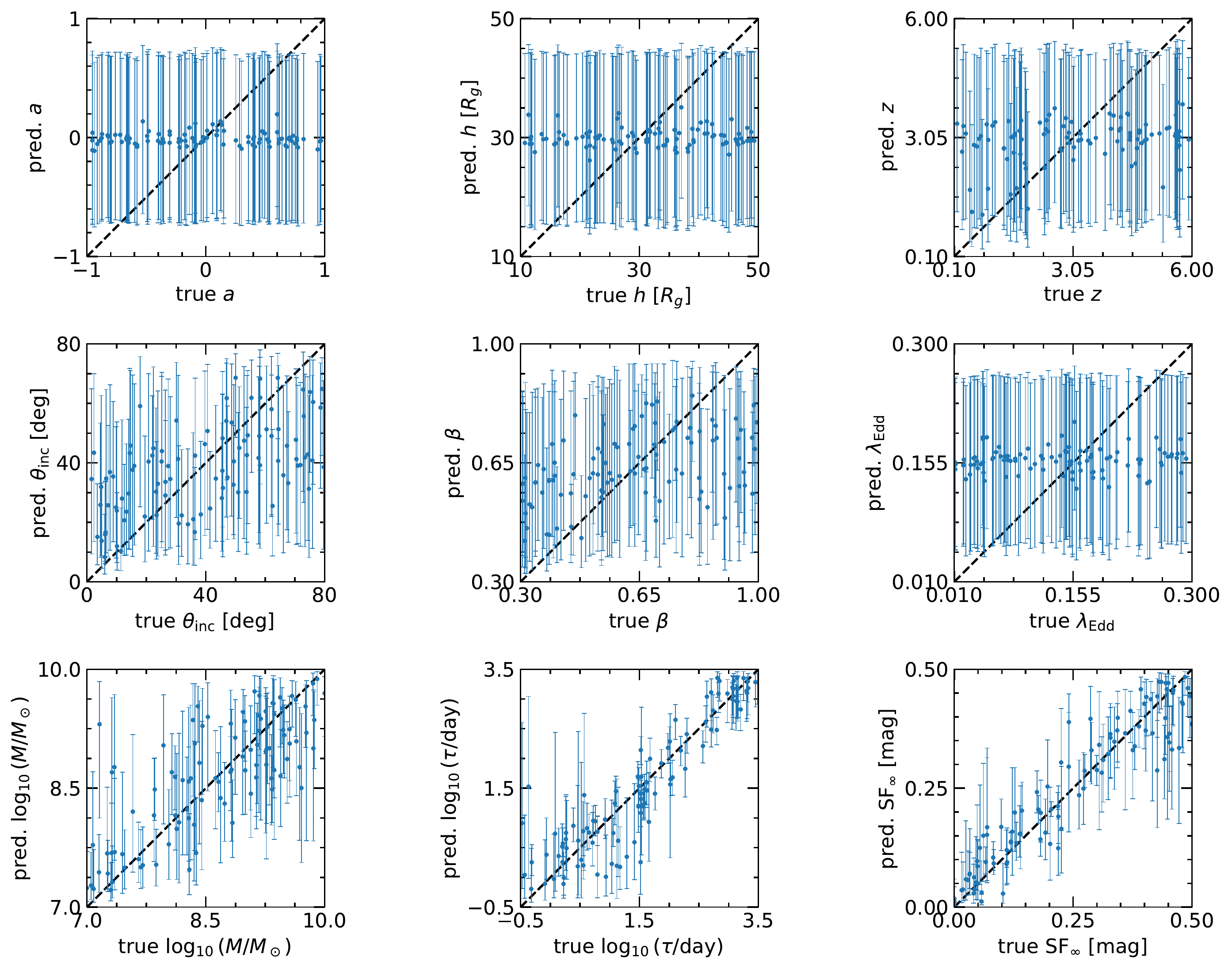}
\caption{Examples of median parameter predictions and $1\sigma$ uncertainties for 100 random light curves in our test set. The black dashed lines along the diagonal represent perfect predictions.}
\label{fig:example_median_pred_with_unc}
\end{figure*}

Figure~\ref{fig:example_median_pred_with_unc} gives 100 random examples of parameter predictions with uncertainties compared to the true values. 

\section{Example of Parameter Posterior} \label{sec:appendix_param_posterior}

\begin{figure*}
\centering
\includegraphics[width=1.0\textwidth]{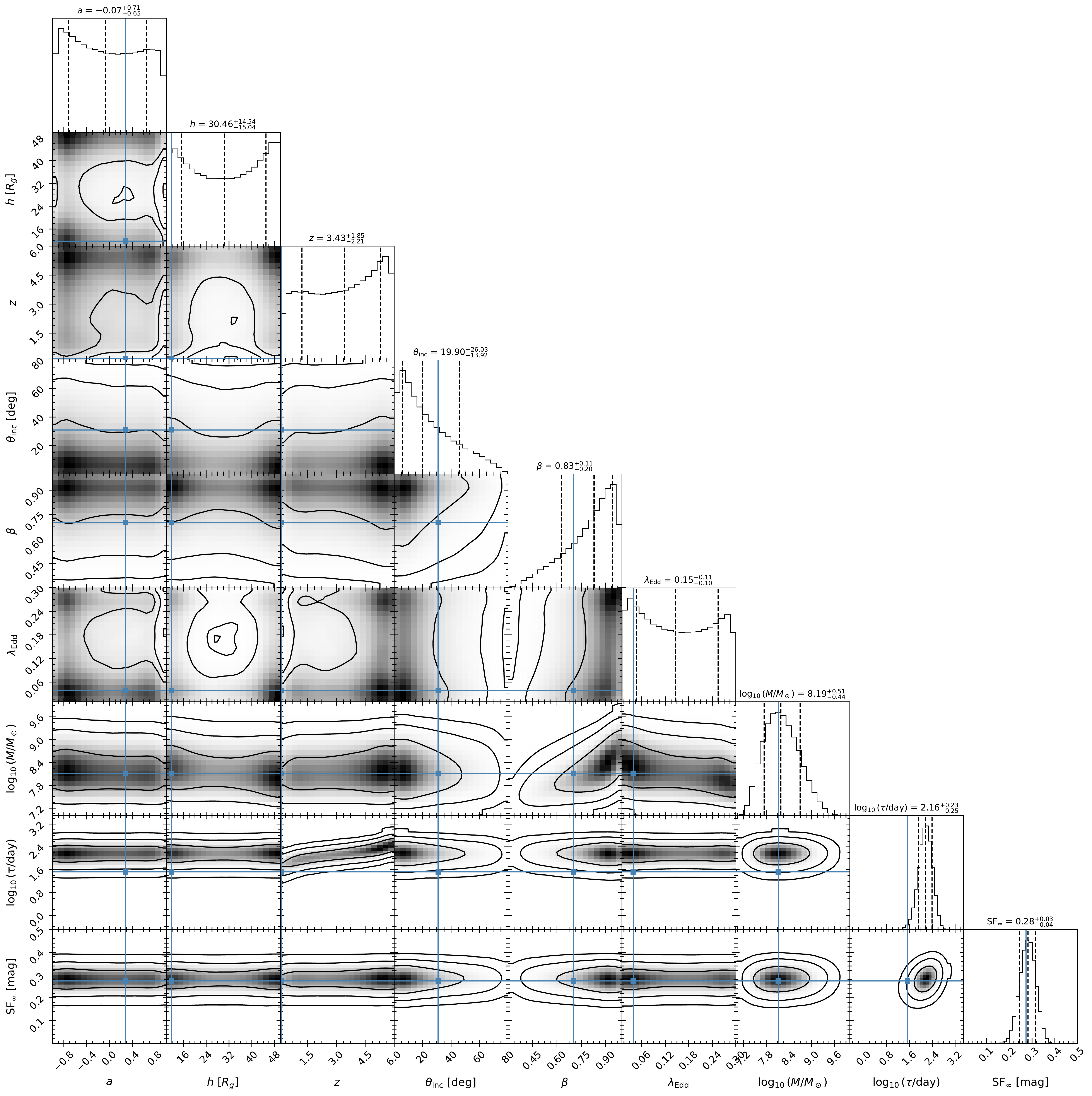}
\caption{Example of a posterior on the parameters for the same test light curve shown in Figure~\ref{fig:recovery}. The diagonal elements display marginal distributions with the median and 1$\sigma$ levels indicated by dashed lines. The central elements depict 1$\sigma$, 2$\sigma$, and 3$\sigma$ contour levels. The true parameter values are overlaid in blue.
}
\label{fig:corner}
\end{figure*}

Figure~\ref{fig:corner} shows an example of a predicted parameter posterior.

\bibliography{bib}{}
\bibliographystyle{aasjournal}

\end{document}